\documentclass[12pt]{article}
\usepackage{a4wide, amsmath, latexsym, epsfig,amsfonts,
  psfrag,graphicx, rotating, fancyhdr, tabularx, afterpage,booktabs}
\usepackage{subfigure}
\usepackage{amssymb}
\usepackage{setspace}

\newcommand\ba{\begin{eqnarray}}
\newcommand\ea{\end{eqnarray}}

\begin{document}
\begin{titlepage}

\begin{flushright}
{{IFJPAN-IV-2012-15}}
\end{flushright}

\vspace{0.2cm} 
\begin{center}

{\Huge \bf Status of TAUOLA and related projects}
\end{center}
\vspace*{5mm}

\begin{center}
   {\bf Z. W\c{a}s$^{a,b}$}\\
       {\em $^a$  Institute of Nuclear Physics, PAN,
        Krak\'ow, ul. Radzikowskiego 152, Poland}\\
{\em $^b$ CERN PH-TH, CH-1211 Geneva 23, Switzerland.}
\end{center}
\vspace{.1 cm}
\begin{center}
{\bf   ABSTRACT  }
\end{center}

Status of   new hadronic currents for  
  $\tau$ lepton decay Monte Carlo generator {\tt TAUOLA} was revieved in other talks of the
conference. Efforts on
comparison with BaBar and Belle
collaboration data were carefully discussed. Also use of the program in phenomenology of $W$ decays
measured by ATLAS collaboration was presented in these talks   as well. 

That is why, in my talk, I will concentrate on
 other aspects of our work 
necessary  for development of  $\tau$ lepton Monte Carlo programs and their phenomenological use.

Presented results illustrate the status of the projects performed in collaboration with 
Swagato Banerjee, Zofia Czyczula, Nadia Davidson, Jan Kalinowski, Wojciech Kotlarski
Tomasz Przedzi\'nski, Olga Shekhovtsova, El\.zbieta Richter-Wa\c{}s, Pablo Roig, Jakub Zaremba, Qingjun Xu and others.

\vskip 1 cm
\begin{center}

{\it \small Presented at International Workshop on Tau Lepton Physics, \\ TAU12 Nagoya, Japan, September, 2012}
\end{center}

 \vspace{4cm}
\begin{flushleft}
{   IFJPAN-IV-2012-15 \\  
 December, 2012}
\end{flushleft}
\end{titlepage}

% \linenumbers

%%%%%%%%%%%%%
\section{Introduction}

%{\tt /home/wasm/y2004/transparencje/kontrybucja /TAU04-ZWas }
%%%%%%%%%%%%%%%%%%%%%%%%%%%%%%%%%%%%%%%%%%%%%%%%%%%%%%%%%%%%%%%%%%%%%%%%%%%%%%%%%%%%%%%%%%%%%
The {\tt TAUOLA} package
\cite{Jadach:1990mz,Jezabek:1991qp,Jadach:1993hs,Golonka:2003xt} for simulation
of $\tau$-lepton decays and
{\tt PHOTOS} \cite{Barberio:1990ms,Barberio:1994qi,Golonka:2005pn} for simulation of QED radiative corrections
in decays, are computing
projects with a rather long history. Written and maintained by
well-defined (main) authors, they nonetheless migrated into a wide range
of applications where they became ingredients of
complicated simulation chains. As a consequence, a large number of
different versions are presently in use. Those modifications, especially in case of
{\tt TAUOLA}, are   valuable from the physics point of view, even though they
 often did not find the place in the distributed versions of
the program.
From the algorithmic point of view, versions may
differ only in  details, but they incorporate many specific results from distinct
$\tau$-lepton measurements or phenomenological projects.
Such versions were mainly maintained (and will remain so)
by the experiments taking precision data on $\tau$ leptons.
Interesting from the physics point of view changes are still
developed in {\tt FORTRAN}.
That is why, for convenience of such partners, part of the
{\tt TAUOLA} should remain in {\tt FORTRAN} for a few forthcoming years.

Many new applications were developed in {\tt C++},  often requiring
a program interface to other packages  (e.g., generating events for LHC, LC,
Belle or BaBar physics processes). For the manipulation of matrix element, techniques of 
re-weighting events were further developed. This required attention on numerical stability issues.

The program structure did not change significantly 
since previous
$\tau$ conference \cite{Was:2011tv}.
Let us concentrate on physics extentions and novel applications.
We will only mention work on new hadronic currents based on the Resonance Chiral approach.
This topic was covered in other talks of the conference \cite{Olga_proceed,Pablo_proceed,Ian_proceed}. Important results 
are already obtained, but sufficiently good agreement with the experimental data is not yet achieved.
New currents are not integrated into main distribution tar-balls for {\tt FORTRAN} and {\tt C++} applications.
Further work is on-going, weighted event 
techniques useful  for fits are used.
Analyses of high precision,
high-statistics  data from Belle and BaBar are expected to profit from these
solutions. 
Other aspects of the project such as interfaces
for applications based on {\tt HepMC} \cite{Dobbs:2001ck} event record
or new tests and weighting  algorithms for spin effects
in production processes should be  mentioned as well. In this context numerical stability 
of solutions used in re-weighting events stored in datafiles is of importance.

Our presentation is organized as follows:
Section 2  is devoted to the discussion  of optional
weights in {\tt TAUOLA} and their use for fits to experimental data.
In section 3 we concentrate on   {\tt PHOTOS} Monte Carlo for
radiative corrections in decays.
Section 4 is devoted to new interfaces of {\tt TAUOLA} and {\tt PHOTOS} based
on  {\tt HepMC} and written in {\tt C++}. Work on interface to
genuine weak corrections, transverse spin effects and new tests
and implementation
bremsstrahlung kernels will be presented as well.
Comments on  changes in {\tt MC-TESTER}; the program designed for
semi-automatic comparisons of simulation samples originating from
different programs and heavily used in our projects are also given.
Section 5 Summary closes the talk.

Because of the limited space of the contribution,
 some results  will not be presented in the
proceedings. They find their place in
publications, prepared with coauthors listed in the Abstract.
For these works,  the present paper may serve as an advertisement.

%%%%%%%%%%%%%%%%%%%%%%%%%%%%%%%%%%%%%%%%%%%%%%%%%%%%%%%%%%%%%%%%%%%%%%%%%%%%%%%%%%%%%%%%%%%%%
\section{ Approach of Resonance Chiral lagrangians and   {\tt TAUOLA} Monte Carlo}
%%%%%%%%%%%%%%%%%%%%%%%%%%%%%%%%%%%%%%%%%%%%%%%%%%%%%%%%%%%%%%%%%%%%%%%%%%%%%%%%%%%%%%%%%%%%%

In other talks \cite{Olga_proceed,Pablo_proceed,Ian_proceed} of the conference, Resonance Chiral Lagrangian
approach was used for calculations of new hadronic currents
to be installed  {\tt TAUOLA}. That is why, we do not need to repeat its description here. 
In Ref.~\cite{Shekhovtsova:2012ra}
implementation of those currents is documented in a great detail.

Physics of $\tau$ lepton decays requires sophisticated strategies for the
confrontation of phenomenological models with experimental data. On one hand,
high-statistics experimental samples are collected, and the obtained precision is
high, on the other hand, there is a significant cross-contamination between distinct
$\tau$ decay channels. Starting from  a certain precision  level all channels
need to be analyzed simultaneously. Change of parameterization for one channel
contributing  to the background to another one may be important for the fit of
its currents. This situation leads to a complex configuration where a multitude of parameters (and models)
needs to be simultaneously confronted with a multitude of observables.
One has to keep in mind that the models used to obtain distributions in
 the fits may require refinements or even substantial rebuilds as a consequence
of comparison with the data. The topic was covered in detail in the $\tau$ Section of Ref.~\cite{Actis:2010gg}. At present our comparison with the data still do not require such refined methods.

We enable  calculation for each generated event (separately for 
decay of $\tau^+$ and/or $\tau^-$) alternative weights; the ratios
of the matrix element squared obtained with new currents,
and the one actually used in generation. Then, the vector of weights can be obtained
and used in fits.
We have checked that such a solution not only can be easily installed into
{\tt TAUOLA} as a stand-alone generator, but it can also be incorporated into
the simulation frameworks of Belle and BaBar collaborations.
The weights can be calculated after the simulation of detector response is
completed. Only then choice of parameters for the hadronic currents has to be
performed and the fits completed. This idea was also behind  {\tt TauSpinner}
for LHC applications, described in Section 4.

%%%%%%%%%%%%%%%%%%%%%%%%%%%%%%%%%%%%%%%%%%%%%%%%%%%%%%%%%%%%%%%%%%%%%%%%%%%%%%%%%%%%%%%%%%%%%
\section{{\tt PHOTOS} Monte Carlo for bremsstrahlung and its systematic uncertainties}
\def\CCol{{\tt SANC}}
%%%%%%%%%%%%%%%%%%%%%%%%%%%%%%%%%%%%%%%%%%%%%%%%%%%%%%%%%%%%%%%%%%%%%%%%%%%%%%%%%%%%%%%%%%%%%
Thanks to exponentiation properties and factorization, the bulk of the final state
QED bremsstrahlung can be described in a universal way.
However, the
kinematic configurations caused by QED bremsstrahlung are affecting
in an  important way
signal/background separation. It may affect selection criteria and background
contaminations in quite complex and unexpected ways.
In many applications, not only in $\tau$ decays,
such bremsstrahlung corrections are
generated with the help of the  {\tt PHOTOS} Monte Carlo. That is why it is of importance to
review the precision of this program as documented in
Refs.~\cite{Barberio:1990ms,Barberio:1994qi,Golonka:2005pn}.
For the {\tt C++} applications, the version of the program is available now.
It is documented in Ref.~\cite{Davidson:2010ew}.

In {\tt C++} applications, the complete first-order matrix elements  for the
 two-body decays of the $Z$ \cite{Golonka:2006tw} and $W$ \cite{Nanava:2009vg}
decays into a lepton pair are now available.
Kernels with complete matrix elements, for the decays of
scalar $B$ mesons  into a pair of scalars  \cite{Nanava:2006vv} are
available for the {\tt C++} users as well.
For  $K \to l \nu \pi$
and  for $\gamma^* \to \pi^+\pi^-$ decays \cite{Nanava:2009vg,Xu:2012px}
matrix element based kernels are still available for tests only.
Properly  oriented reference frames are needed in those cases.
It will be rather easy to
integrate those NLO kernels  into the main version of the program,
 because of better control of the
decay particle rest frame than in the {\tt FORTRAN} interface.

In all of these cases the universal kernel of {\tt PHOTOS} is replaced with the
one matching an exact first-order matrix element. In this way terms necessary
for the NLO/NLL precision 
level are implemented\footnote{Note that here the LL (NLL) denotes
 collinear logarithms (or in case of differential
predictions terms integrating into such logarithms).
 The logarithms of soft singularities are taken into
account to all orders. This is resulting from mechanisms of exclusive
exponentiation \cite{Jadach:2000ir} of QED.
The algorithm used in {\tt PHOTOS} Monte Carlo is compatible with exclusive
exponentiation. Note that our
 LL/NLL precision level would even read  as  respectively   NLL/NNNLL
 level in some naming conventions of QCD.
}.
A discussion relevant for control of program systematic uncertainty in $\tau \to \pi \nu$ decay can be found in
Ref.~\cite{Guo:2010ny}.

The algorithm covers the full multiphoton
phase-space and becomes  exact in the soft limit.
This is rather unusual for  NLL compatible algorithms. One should not forget
that {\tt PHOTOS} generates weight-one events, and does not exploit any
phase space ordering. There is a full phase space overlap between the one
where a hard
matrix element is used and the one for iterated photon emission.
All interference effects (between consecutive emissions and emissions from
distinct charged lines) are implemented with the help of internal weights.

The results of all tests of {\tt PHOTOS} with a NLO kernel confirm
 sub-permille precision level.
This is very encouraging, and points to the possible extension of the
approach outside of  QED (scalar QED). In particular, to the domain of
QCD or to QED when  phenomenological form factors for interactions
of photons need
 to be used. For that work  to be completed, spin amplitudes need
to be studied. Let us point to Ref.~\cite{vanHameren:2008dy}
as an example.

New tests of {\tt PHOTOS} are available from Ref.~\cite{Photos_tests}.
In those tests, in particular,
results from the second-order matrix element calculations
embedded in  {\tt KKMC} \cite{Jadach:1999vf} Monte Carlo are used in case of $Z$ decay.
For $W$ decays comparisons with electroweak calculations of
Refs.~\cite{Andonov:2008ga,Andonov:2004hi} are shown.

%%%%%%%%%%%%%%%%%%%%%%%%%%%%%%%%%%%%%%%%%%%%%%%%%%%%%%%%%%%%%%%%%%%%%%%%%%%%%%%%%%%%%%%%%%%%%
\section{  {\tt TAUOLA universal interface} and {\tt PHOTOS} interface in {\tt C++}}
%%%%%%%%%%%%%%%%%%%%%%%%%%%%%%%%%%%%%%%%%%%%%%%%%%%%%%%%%%%%%%%%%%%%%%%%%%%%%%%%%%%%%%%%%%%%%

\begin{figure}
\begin{center}
\setlength{\unitlength}{0.5 mm}
\begin{picture}(35,80)
%%%\put( 0,0){\framebox( 60,50){ }}
\put( -65,-45){\makebox(0,0)[lb]{\epsfig{file=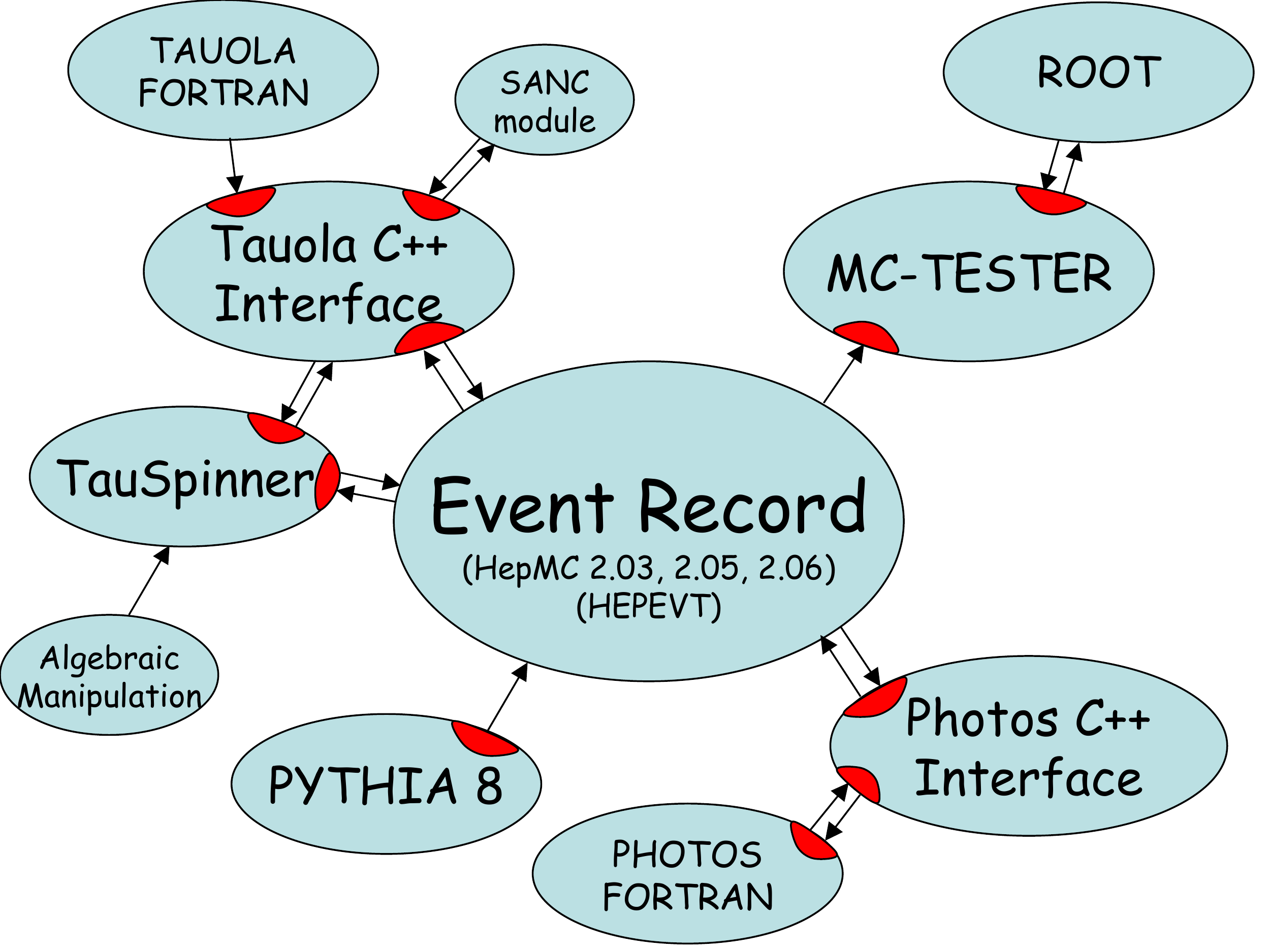,width=70mm,height=65mm}}}
\end{picture}
\end{center}
\vskip 1.5 cm
\caption{\small \it  Scheme of Monte Carlo simulation system with 
communication based on event record. Each segment feature contribution 
from different people, may be coded in distinct programinng language
and/or be developped with the help of algebraic manipulation systems.
 } \label{Relations}
\end{figure}

In the development of packages such as {\tt TAUOLA} or {\tt PHOTOS}, questions
of tests and appropriate relations to users' applications are essential for
their
usefulness. In fact, user applications may be much larger in size and
human efforts than the programs discussed here.
Good example of such `user applications' are complete environments to simulate
physics process and control detector response at the same time.
Distributions of final state particles are not always of direct interest.
Often properties of intermediate states, such as a spin state of $\tau$-lepton,
coupling constants or masses of intermediate heavy particles are
of prime interest.
As a consequence, it is useful that such intermediate state properties are
under direct control of the experimental user and can be manipulated
to understand detector responses.
Our programs  worked well   with {\tt FORTRAN} applications  where {\tt HEPEVT} event record
is used.  For the  {\tt C++}  {\tt HepMC} \cite{Dobbs:2001ck} case,
interfaces were  rewritten, both for
{\tt TAUOLA} \cite{Davidson:2010rw} and for {\tt PHOTOS} \cite{Davidson:2010ew}.
The interfaces and as a consequence the programs themselves
were enriched; for  {\tt PHOTOS} new Matrix element kernels
are  available; for {\tt TAUOLA} interface,
 a complete (not  longitudinal only) spin correlations
 are available
for $Z/\gamma^*$ decay.
Electroweak corrections taken from
Refs.~\cite{Andonov:2008ga,Andonov:2004hi} are also used.
For the scheme
of programs communications see Fig.~\ref{Relations}.
In this spirit an algorithm of {\tt TauSpinner} \cite{Czyczula:2012ny} to study detector response to spin effects
in $Z, W$ and $H$ decays, was developed. Recently {\tt TauSpinner} was
enriched \cite{Banerjee:2012ez} with the option to study effects of New Physics, such as
effects of spin-2 states in $\tau^+\tau^-$ pairs produced at LHC. 
Modular organization opens ways for further efficient algorithms to understand
detector systematics, but at the same time responsibility to control software
precision must be shared by the user.
Automated   tests of
{\tt MC-TESTER} were prepared~\cite{Golonka:2002rz}.
New functionalities
were introduced into the testing package \cite{Davidson:2008ma}. In particular, it works now with the
{\tt HepMC} event record, the  standard of {\tt C++} programs,
spectrum of available tests is enriched and events stored on datafiles are easier to test.

%%%%%%%%%%%%%%%%%%%%%%%%%%%%%%%%%%%%%%%%%%%%%%%%%%%%%%%%%%%%%%%%%%%%%%%%%%%%
%%%%%%%%%%%%%%%%%%%%%%%%%%%%%%%%%%%%%%%%%%%%%%%%%%%%%%%%%%%%%%%%%%%%%%%%%%%%

The program is available
through the LHC Computing Grid ({\tt LCG}) Project. See
{\tt GENSER} webpage, Ref.~\cite{Kirsanov:2008zz}, for details.
 This is the case for {\tt TAUOLA}  {\tt C++} and for
{\tt PHOTOS}  {\tt C++}  as well.
The {\tt FORTRAN} predecessors are
available in this way too.

%%%%%%%%%%%%%%%%%%%%%%%%%%%%%%%%%%%%%%%%%%%%%%%%%%%%%%%%%%%%%%%%%%%%%%%%%%%%
%%%%%%%%%%%%%%%%%%%%%%%%%%%%%%%%%%%%%%%%%%%%%%%%%%%%%%%%%%%%%%%%%%%%%%%%%%%%
\section{Summary and future possibilities}

Versions of the hadronic currents available for the {\tt TAUOLA} library
until now, are all based on old models and experimental data of 90's.
The implementation of   new currents, based on the Resonance Chiral Lagrangian approach
 is now prepared
and tested from the technical side. Methods for efficient confrontation
with the experimental data are prepared as well.
Once comparison with Belle and BaBar data is successfully completed,
new parameterizations will be straightforward for use in
a spectrum of applications
in {\tt FORTRAN} or {\tt C++} environments.

The status of
associated projects: {\tt TAUOLA universal interface } and {\tt MC-TESTER}
was reviewed. Also
the high-precision version of  {\tt PHOTOS} for radiative corrections in
decays, was
presented. All these programs are available now for {\tt C++} applications
thanks to the {\tt HepMC} interfaces.

New results for  {\tt PHOTOS}   were mentioned.
For the leptonic $Z$ and $W$ decays  the complete next-to-leading collinear logarithms
effects can now be  simulated in {\tt C++} applications.
However, in  most cases these
effects are not important, leaving the standard  version
 sufficient.
 Thanks to this work  the path for  fits to the data of
electromagnetic form factors
 is opened~\cite{Xu:2012px}, e.g. in the  case of $K_{l3}$ decays.

 The presentation of the {\tt TAUOLA} general-purpose interface
in {\tt C++} was given. It is more refined
than the {\tt FORTRAN} predecessor. Electroweak corrections can be used
in calculation of complete spin correlations in $Z/\gamma^*$ mediated
processes.  An algorithm for study of detector responses to spin effects in
$Z$, $W$ and $H$ decays was shown.

The present version of {\tt MC-TESTER} is stable now.
It works with {\tt C++} event record {\tt HepMC}   and enables
user defined tests in experiments' software environments.  We used the tool
regularly in  our projects.

\section{Acknowledgement}

This research was supported in part by the funds of Polish National Science Centre under decision DEC-2012/04/M/ST2/00240 and DEC-2011/03/B/ST2/00107 and   by the Polish Government grant NN202127937 (years 2009-2011).


%% References

%\nocite{*}
%\bibliographystyle{elsarticle-num}
%\bibliography{martin}






% \end{thebibliography}

\providecommand{\href}[2]{#2}\end{document}